\documentclass[nofootinbib,twocolumn,preprintnumbers,prl]{revtex4-1}
\pdfoutput=1
\usepackage{amsmath,amsthm,amssymb,multirow,psfrag}
\usepackage{epsfig}
\usepackage{color}
\usepackage[utf8]{inputenc}

\graphicspath{{./Figures/}}

\begin{document}

\def\lsim{\mathrel{\rlap{\lower4pt\hbox{\hskip1pt$\sim$}}
  \raise1pt\hbox{$<$}}}
\def\gsim{\mathrel{\rlap{\lower4pt\hbox{\hskip1pt$\sim$}}
  \raise1pt\hbox{$>$}}}
\newcommand{\vev}[1]{ \left\langle {#1} \right\rangle }
\newcommand{\bra}[1]{ \langle {#1} | }
\newcommand{\ket}[1]{ | {#1} \rangle }
\newcommand{\ev}{ {\rm eV} }
\newcommand{\kev}{{\rm keV}}
\newcommand{\mev}{{\rm MeV}}
\newcommand{\tev}{{\rm TeV}}
\newcommand{\mpl}{$M_{Pl}$}
\newcommand{\mw}{$M_{W}$}
\newcommand{\Ft}{F_{T}}
\newcommand{\Zparity}{\mathbb{Z}_2}
\newcommand{\BLambda}{\boldsymbol{\lambda}}
\newcommand{\met}{\;\not\!\!\!{E}_T}
\newcommand{\beq}{\begin{equation}}
\newcommand{\eeq}{\end{equation}}
\newcommand{\bea}{\begin{eqnarray}}
\newcommand{\eea}{\end{eqnarray}}
\newcommand{\nn}{\nonumber}
\newcommand{\gev}{{\mathrm GeV}}
\newcommand{\hc}{\mathrm{h.c.}}
\newcommand{\eps}{\epsilon}
\newcommand{\bwt}{\begin{widetext}}
\newcommand{\ewt}{\end{widetext}}
\newcommand{\draftnote}[1]{{\bf\color{blue} #1}}

\newcommand{\cO}{{\cal O}}
\newcommand{\cL}{{\cal L}}
\newcommand{\cM}{{\cal M}}

\newcommand{\fref}[1]{Fig.~\ref{fig:#1}} 
\newcommand{\eref}[1]{Eq.~\eqref{eq:#1}} 
\newcommand{\aref}[1]{Appendix~\ref{app:#1}}
\newcommand{\sref}[1]{Section~\ref{sec:#1}}
\newcommand{\tref}[1]{Table~\ref{tab:#1}}

\title{\Large{{\bf The Virtual Diphoton Excess}}}
\author{{\bf{Daniel Stolarski$\,^{a,b}$~and~Roberto Vega-Morales$\,^{c}$}}}

\affiliation{
$^a$Theoretical Physics Department, CERN, Geneva, Switzerland \\
$^b$Ottawa-Carleton Institute for Physics, Carleton University, 1125 Colonel By Drive, Ottawa, Ontario K1S 5B6, Canada \\
$^{c}$Departamento de F\'{i}sica Te\'{o}rica y del Cosmos and CAFPE,
Universidad de Granada,\\ Campus de Fuentenueva, E-18071 Granada, Spain
}

\preprint{CERN-TH-2016-005}

\email{
daniel.stolarski@cern.ch\\
rvegamorales@ugr.es\\}

\begin{abstract}
%
%
%
Interpreting the excesses around 750 GeV in the diphoton spectra to be the signal of a new heavy scalar $\varphi$ decaying to photons, we point out the possibility of looking for correlated signals with virtual photons.~In particular, we emphasize that the effective operator that generates the $\varphi \to \gamma\gamma$ decay will also generate decays of $\varphi \to 2\ell \gamma$ ($2\ell \equiv 2e, 2\mu$) and $\varphi \to 4\ell$ ($4\ell \equiv 2e2\mu, 4e, 4\mu$) independently of the $\varphi$ couplings to $Z\gamma$ and $ZZ$.~Depending on the relative sizes of these effective couplings, we show that the virtual diphoton component can make up a sizable, and sometimes dominant, contribution to the total $\varphi \to 2\ell \gamma$ and $\varphi \to 4\ell$ partial widths.~We also discuss modifications to current experimental cuts in order to maximize the sensitivity to these virtual photon effects.~Finally, we briefly comment on prospects for channels involving other Standard Model fermions as well as more exotic decay possibilities of the putative resonance.  
\end{abstract}

\maketitle

\section{Introduction} \label{sec:intro} 
There has been tremendous interest in the excesses recently reported by both ATLAS~\cite{ATLASdiphoton} and CMS~\cite{CMS:2015dxe} in the diphoton spectrum around 750 GeV.~If this is a sign of a new resonance, the simplest explanation for the decay is through the photon field strength or dual field strength tensor.~For concreteness we will consider the dual field strength case via the dimension five operator
\bea
\frac{c_{\gamma\gamma}}{4\Lambda} \, \varphi\, F_{\mu\nu}\widetilde{F}^{\mu\nu}, 
\label{eq:photon-tensor}
\eea
where $F_{\mu\nu}=\partial_\mu A_\nu - \partial_\nu A_\mu$ and $\widetilde{F}^{\mu\nu}=\frac{1}{2}\epsilon^{\mu\nu\rho\sigma}F_{\rho\sigma}$.~We take $\Lambda$ to be some new mass scale associated with this operator that will cancel in all the ratios we will consider.~Our choice of operator in~\eref{photon-tensor} implies the new resonance $\varphi$ is a parity odd scalar, but our considerations largely apply if it turns out to be a parity even or CP violating scalar as well as a spin 2 resonance.

Assuming electroweak $SU(2)\times U(1)$ gauge symmetry holds in the UV, the operator in~\eref{photon-tensor} must descend from a linear combination of the operators~\cite{Low:2015qho}:
\bea
\frac{c_{W}}{4\Lambda} \, \varphi \, W^a_{\mu\nu}\widetilde{W}^{a\mu\nu} \;\;\;\; {\rm and} \;\;\;\; 
\frac{c_{B}}{4\Lambda}\, \varphi\, B_{\mu\nu}\widetilde{B}^{\mu\nu} .
\label{eq:SM-tensor}
\eea
As has already been pointed out many times~\cite{Cai:2015bss,Harigaya:2015ezk,Mambrini:2015wyu,Nakai:2015ptz,Knapen:2015dap,Buttazzo:2015txu,Franceschini:2015kwy,Higaki:2015jag,McDermott:2015sck,Ellis:2015oso,Bellazzini:2015nxw,Low:2015qep,Gupta:2015zzs,Molinaro:2015cwg,Cao:2015pto,Matsuzaki:2015che,Kobakhidze:2015ldh,Cox:2015ckc,Curtin:2015jcv,Bian:2015kjt,Ahmed:2015uqt,Falkowski:2015swt,Bai:2015nbs,Benbrik:2015fyz,Alves:2015jgx,Cao:2015twy,Liao:2015tow,deBlas:2015hlv,Belyaev:2015hgo,Altmannshofer:2015xfo,Craig:2015lra,Cao:2015scs,Dev:2015vjd},
these operators will lead to correlated signals in $\varphi$ decays to $Z\gamma$ and $ZZ$, as well as $WW$ if $c_W$ is non-zero.~Searches for diboson resonances have been performed by ATLAS~\cite{Aad:2014fha,Aad:2015ipg,Aad:2015rka} and CMS~\cite{Khachatryan:2014gha}~placing constraints on models which can explain the diphoton resonance.

In this letter, we emphasize that the operator in~\eref{photon-tensor} \textit{alone} is enough to produce $\varphi \to 2f\gamma$ and $\varphi \to 4f$ decays of the $\varphi$ resonance through virtual photons, irrespective of its UV origin.~We examine under which circumstances the virtual photon component makes up a sizable contribution, or even dominates over the $ZZ$ and $Z\gamma$ components, to these three and four body decays, with particular emphasis on the leptonic $\varphi \to 2\ell\gamma~(2\ell \equiv 2e, 2\mu)$ and $\varphi \to 4\ell~(4\ell \equiv 2e2\mu, 4e, 4\mu)$ channels.

We also examine what effects cuts on the lepton invariant masses have on the relative composition of the $\varphi \to 2\ell\gamma$ and $\varphi \to 4\ell$ partial widths.~Should the diphoton excess persist, then knowing the mass of $\varphi$ will allow a search for $\varphi \to 2\ell\gamma$ and $\varphi \to 4\ell$ decays imposing only minimal constraints on any subset of the final states.~We take advantage of this to motivate modifying current experimental searches in the $2\ell\gamma$ and $4\ell$ channels in order to maximize the sensitivity to the virtual diphoton effects.~We also briefly discuss possibilities in the less experimentally clean decays to other SM fermions.

All of the results presented here are obtained by integration of the analytic expressions for the $\varphi \to 2\ell\gamma$ and $\varphi \to 4\ell$ fully differential decay widths obtained in~\cite{Chen:2012jy,Chen:2013ejz,Chen:2014ona} to which we refer the reader for further details.

\section{Decay of $\varphi$ to  $2\ell\gamma$}
\label{sec:llgam}

If there is indeed a new particle decaying to $\gamma\gamma$, then it will also decay to $2\ell\gamma$ via a virtual photon.~The rate of this decay is strongly sensitive to the phase space cuts, particularly on the invariant mass of the lepton pair.~In particular, if an experimental analysis allows lepton pairs with an invariant mass between $M_{\rm low}$ and $M_{\rm high}$, then the ratio of partial widths gives
\bea
\frac{\Gamma(\varphi \rightarrow \gamma^* \gamma 
\rightarrow 2\ell\gamma)}{\Gamma(\varphi \rightarrow \gamma \gamma )}
\approx \frac{4\alpha}{3\pi} 
\log\left(\frac{M_{\rm high}}{M_{\rm low}}\right) .
\label{eq:estimate}
\eea
The factor of $\alpha/\pi$ comes from the additional photon coupling, while the log comes from integrating the photon propagator over the phase space.~From this formula we see that if a search has a narrow invariant mass window around the $Z$ pole, as in the ATLAS search~\cite{Aad:2014fha} which requires $65 < M_{\ell\ell} < 120$ GeV, then the effects from virtual photons will be tiny.~On the other hand, making a search as inclusive as possible will raise the rate from virtual photons even in the absence of contributions from $Z$'s. 

Of course most models that explain the diphoton excess via~\eref{photon-tensor} will also generate the $Z\gamma$ operator 
\bea
\frac{c_{Z\gamma}}{2\Lambda}\, \varphi\, F_{\mu\nu}\widetilde{Z}^{\mu\nu} .
\eea
Naively, the effects from this operator should be parametrically larger than the the $\gamma^* \gamma$ operator since the $Z$ can be produced on shell.~However, the suppression is not nearly so large for two important reasons:
\begin{itemize}
\item The $Z$ coupling to leptons is suppressed relative to that of the photon.
\item Unlike the photon, there is no log enhancement when integrating the region of phase space away from the $Z$ pole. 
\end{itemize}
Therefore, if the phase space cuts are very inclusive, the off-shell photon can be an important effect.

In~\fref{components} we plot the three different contributions to the process $\varphi \rightarrow 2\ell\gamma$ as a function of the ratio of couplings
\bea
\lambda_{Z\gamma} = c_{Z\gamma}/c_{\gamma\gamma} .
\label{eq:lambdazg}
\eea
We have normalized the three components of $\varphi \to 2\ell\gamma$ to the $\varphi \to \gamma\gamma$ partial width so the ratio involving the $\gamma^*\gamma$ component (blue curve) is flat.~We plot these ratios for both ATLAS-like phase space cuts (solid) and for much more inclusive  `Full' cuts\footnote{Note that we have only considered cuts on the lepton invariant masses and not on the lepton $p_T$ or rapidity.~Since the rate is dominated by the pole structure of the vector boson propogators, this simplifications captures qualitatively the features we wish to emphasize in this study.}
with $4~\rm{GeV} < M_{\ell\ell} < 750~\rm{GeV}$ (dashed).~The lower cutoff of $4$~GeV is inspired by studies looking for similar off-shell photon effects involving the Higgs boson at 125~GeV~\cite{Gonzalez-Alonso:2014rla,Chen:2015iha,Chen:2015rha}.~We see that with these relaxed phase space cuts, the $\gamma^\ast\gamma$ component can be a few per cent of the on-shell rate because the log in \eref{estimate} is large, while with current cuts the virtual photon contribution is an order of magnitude smaller.

\begin{figure}[tb]
\includegraphics[width=0.46\textwidth]{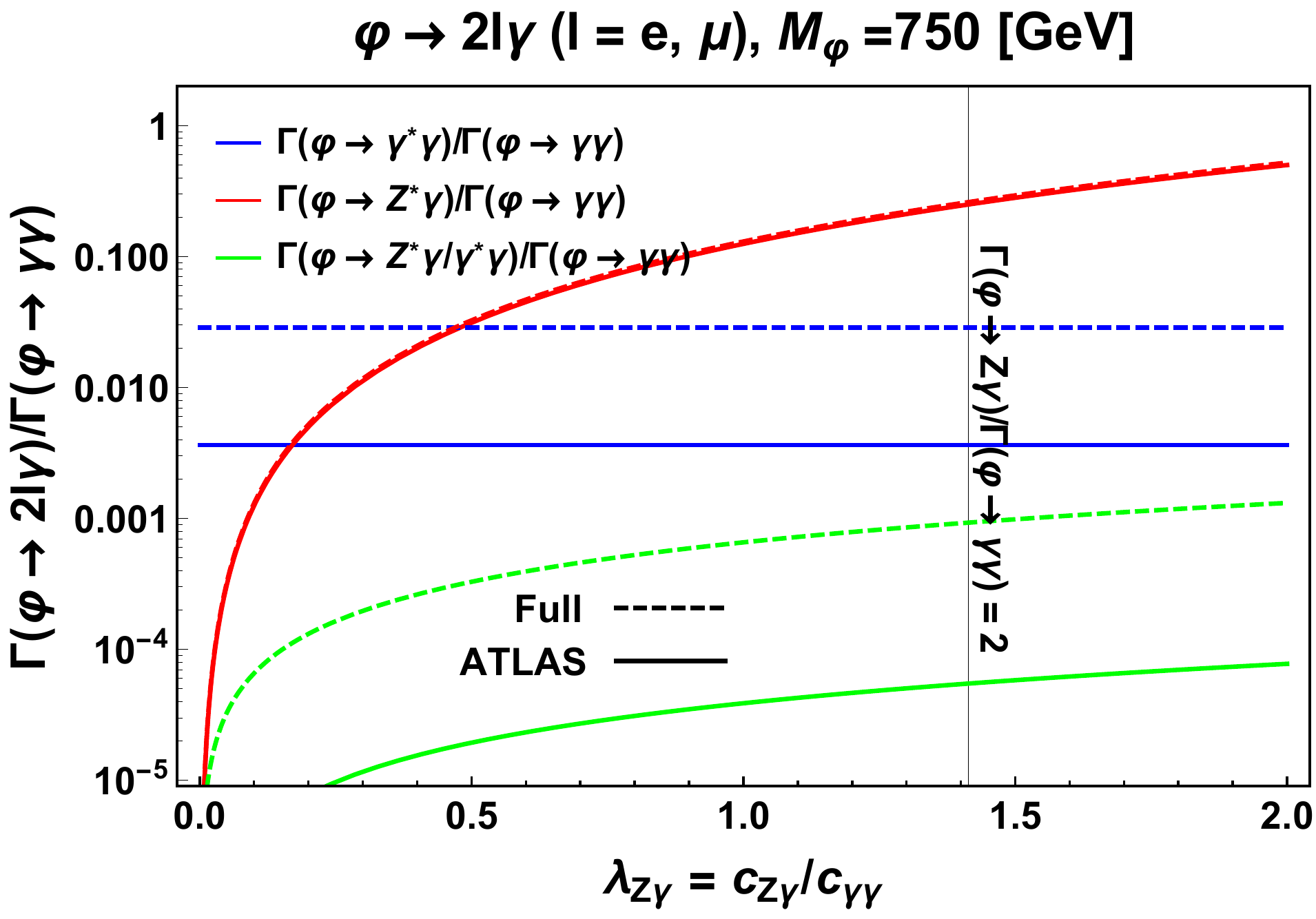}
\caption{The ratio of $\varphi \to 2\ell\gamma$ events that come from the three underlying components:~$Z^*\gamma$ (red),~$\gamma^*\gamma$ (blue), and interference between the two (green), relative to the number of on-shell $\varphi \to \gamma\gamma$ events.~The contributions are shown as a function of $\lambda_{Z\gamma}$, the ratio of couplings defined in~\eref{lambdazg}.~We show ratios for the inclusive (Full) phase space cuts $4 < M_{\ell\ell} < 750$ GeV (dashed) as well as for ATLAS-like cuts with $65~\rm{GeV} < M_{\ell\ell} < 115~\rm{GeV}$ (solid).~The vertical grey line corresponds to Run I limits on $\lambda_{Z\gamma}$ derived from~\cite{Aad:2014fha}. }
\label{fig:components}
\end{figure}

We also see in~\fref{components} that for small $\lambda_{Z\gamma}$, the $\gamma^*\gamma$ component dominates, while for large $\lambda_{Z\gamma}$ the $Z^*\gamma $ component dominates as expected.~Another expected feature is that the contribution from $Z^*\gamma $ is relatively unaltered by these cuts since they both contain the Z-pole.~The interference between the two components is always small, but is significantly enhanced by the more inclusive cuts, making this effect potentially observable with a large number of $\varphi$ decays.~This type of interference also opens up the possibility of observing CP violation in the $\varphi \to 2\ell\gamma$ three body decays as proposed for the Higgs boson~\cite{Chen:2014ona}. 

From the ATLAS 8 TeV search~\cite{Aad:2014fha}, one can bound the cross section into $2\ell\gamma$, although the bound depends on how the cross section scales going from 8 to 13 TeV.~In the case of gluon initiated production, the two body decay $\varphi \to Z\gamma$ is limited to be about twice $\varphi \to \gamma\gamma$ (see for example~\cite{Franceschini:2015kwy}) so we place a grey vertical line to indicate this limit.~The production mechanism could however be photon~\cite{Fichet:2015vvy,Csaki:2015vek} or quark~\cite{Franceschini:2015kwy,Gao:2015igz} initiated, or perhaps some more exotic production mechanism~\cite{Cho:2015nxy,Li:2015jwd,An:2015cgp,Bernon:2015abk,Liu:2015yec}.~Therefore, we show results for even larger values of $\lambda_{Z\gamma}$ due to this uncertainty. 

\begin{figure}[tb]
\includegraphics[width=0.47\textwidth]{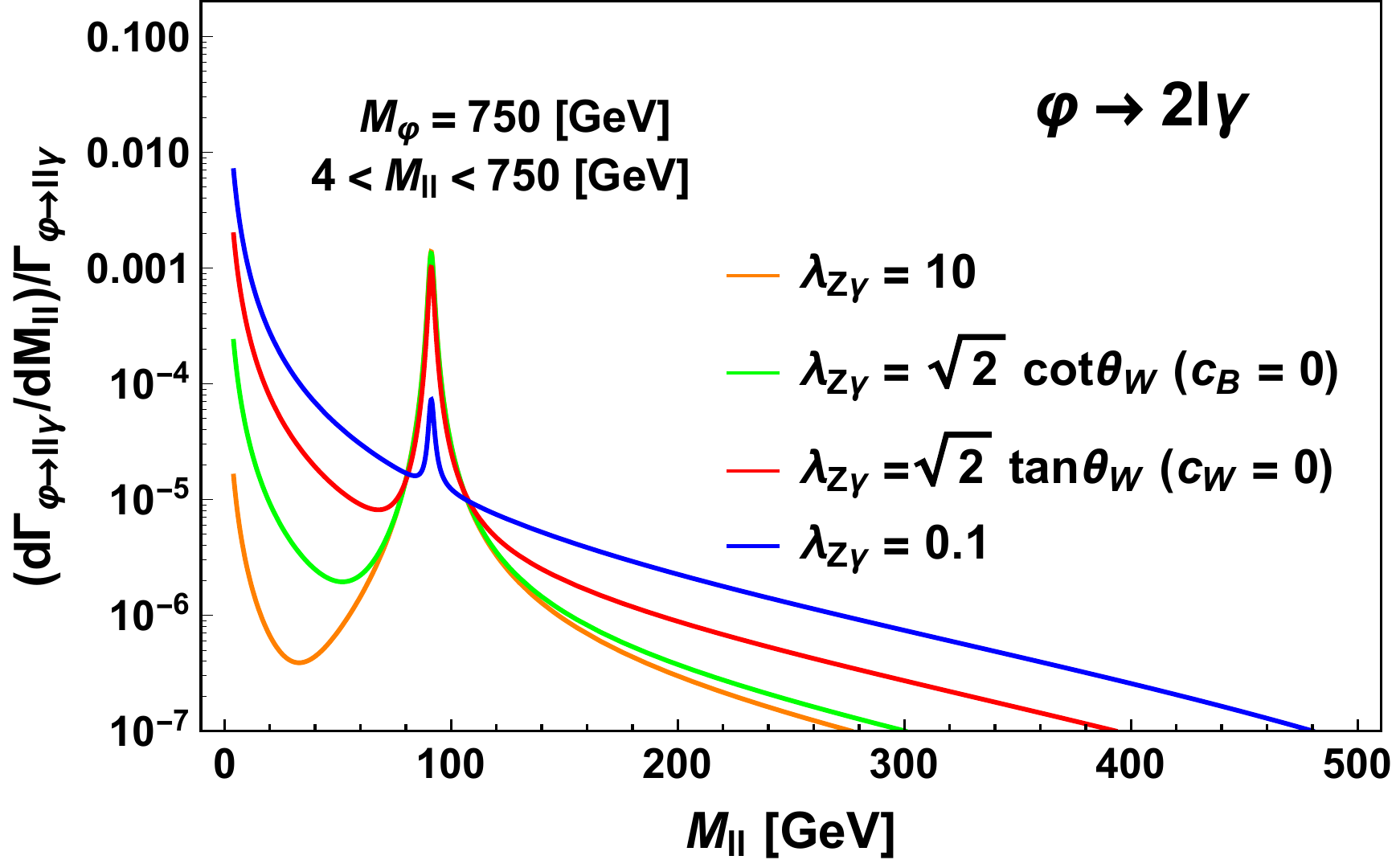}
\caption{Normalized (over $4< M_{\ell\ell}<750$ GeV) lepton pair invariant mass distribution shown for two extreme values 10 (orange) and 0.1 (blue) of the ratio of couplings $\lambda_{Z\gamma}$ defined in Eq.~(\ref{eq:lambdazg}).~We also show the two simple cases of $c_W = 0$ (red) and $c_B = 0$ (green) if the $\varphi Z\gamma$ and $\varphi \gamma\gamma$ operators descend from the $SU(2)\times U(1)$ invariant operators in~\eref{SM-tensor}.~See text for more information.}
\label{fig:Mlldist}
\end{figure}

The central observation of this study is that the invariant mass spectrum of the lepton pair (rather than the full $\ell\ell\gamma$ system) contains significant information on the couplings of the new resonance to gauge bosons.~In Fig.~\ref{fig:Mlldist} we plot the normalized invariant mass distributions for two extreme values (10 and 0.1) of the ratio of couplings $\lambda_{Z\gamma}$ defined in Eq.~(\ref{eq:lambdazg}).~We also show the two simple cases of $c_W = 0$ (red) and $c_B = 0$ (green) using the $SU(2)\times U(1)$ operators in Eq.~\eqref{eq:SM-tensor}. These predict $\lambda_{Z\gamma} = \sqrt{2} \tan\theta_W\approx 0.8$ and $\lambda_{Z\gamma} = \sqrt{2}\cot\theta_W\approx 2.6$ respectively~\cite{Low:2015qho}, where $\theta_W$ is the Weinberg angle.~Unsurprisingly, we see that larger values of $\lambda_{Z\gamma}$ raises the height of the peak around the $Z$ pole, while lower values raises the value at low $M_{\ell\ell}$.~A perhaps more unexpected feature, is that for low values of the ratio there are also more events at high $M_{\ell\ell}$ above the $Z$ peak.~This comes from the fact that the distributions are normalized so the peak is not as large. 

We can exploit the fact that the virtual photon and $Z$ have very different distributions in the invariant mass of the lepton pair to make a crude but very simple measurement of $\lambda_{Z\gamma}$.~The idea is to simply take the fraction of events that have leptons near the $Z$ pole:
\bea
R_Z(\Delta) = \frac{N(M_Z + \Delta > M_{\ell\ell} > M_Z - \Delta) }{{\rm total \; number \; of \; events}},
\label{eq:Rz}
\eea
where the total number of events is defined by the inclusive phase space cuts $4~\rm{GeV} < M_{\ell\ell} < 750$~GeV.~As can be seen in~\fref{zgamlambda}, $R_Z$ is strongly dependent on $\lambda_{Z\gamma}$.~We plot various different values of the mass window $\Delta$, and we see that for $\lambda_{Z\gamma} \lesssim 0.7$, the slope of the curve is large and this variable becomes quite sensitive.~For larger couplings, the virtual photon contribution to this channel becomes subdominant and this observable becomes less sensitive.~In this case, however, the total rate of $\varphi \to 2\ell\gamma$ events will be larger so a more statistically precise measurement will be possible.
\begin{figure}[tb]
\includegraphics[width=.45\textwidth]{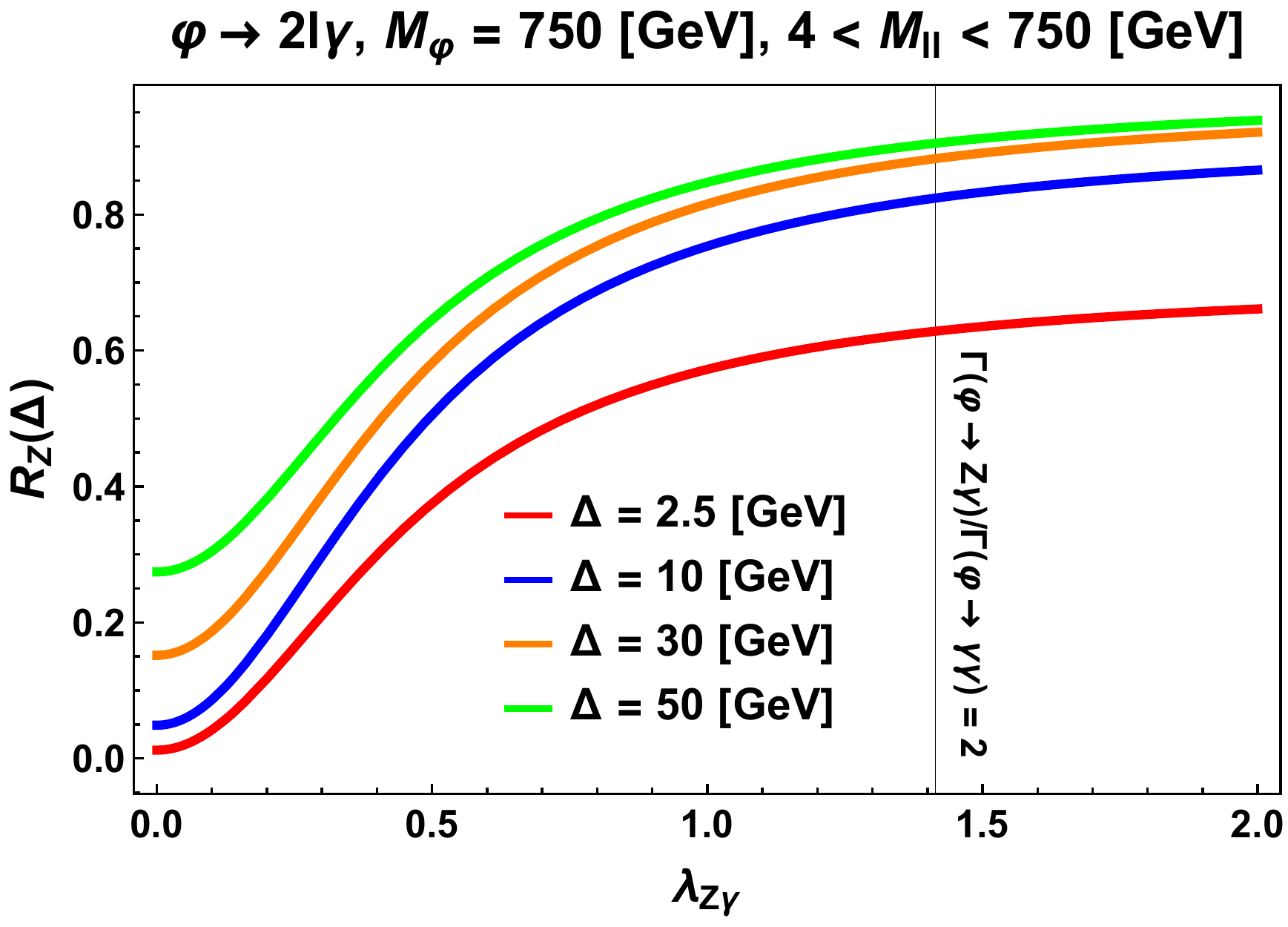}
\caption{$R_Z(\Delta)$, the fraction of events near the $Z$ pole defined in~\eref{Rz} as a function of $\lambda_{Z\gamma}$, the ratio of couplings defined in~\eref{lambdazg}.~We plot $\Delta = 2.5,\;10,\;30,\;50$ GeV going from bottom to top.~The total phase space is defined via the cuts $4~\rm{GeV} < M_{\ell\ell} < 750$ GeV shown at the top.~Again, we also show the limit (vertical line) from $Z\gamma$ searches~\cite{Aad:2014fha} at 8 TeV.}
\label{fig:zgamlambda}
\end{figure}

One could imagine varying $\Delta$ in an experimental analysis to get more information about this coupling ratio.~Taking this to the extreme and using the full phase space information contained in the differential mass distribution event by event would allow for even better measurements.~Of course using a so-called matrix element method where the likelihood is constructed from the fully differential decay width using all observables in $\varphi \to 2\ell\gamma$ uses the maximum amount of information.~Furthermore, at 750~GeV these kinematic observables may be more discriminating than was found for a 125~GeV Higgs boson~\cite{Gainer:2011aa} decaying to $2\ell\gamma$.~However, we leave a fully differential analysis utilizing all observables in $\varphi \to 2\ell\gamma$ using the framework of~\cite{Chen:2012jy,Chen:2013ejz,Chen:2014ona,Chen:2014pia} to ongoing work~\cite{followup}.

Finally, we briefly comment on backgrounds.~The dominant background around 750 GeV in the current search~\cite{Aad:2014fha} comes from genuine $2\ell\gamma$, while a jet faking a photon is the second most important but highly subdominant.~The dominant background has been calculated very precisely in both the $q\bar{q}$ and $gg$ initial states~\cite{Ametller:1985di,vanderBij:1988fb,Ohnemus:1992jn,Baur:1997kz,DeFlorian:2000sg,Adamson:2002rm,Hollik:2004tm,Accomando:2005ra,Campbell:2011bn,Hamilton:2012np,Grazzini:2013bna,Barze:2014zba,Denner:2014bna,Grazzini:2015nwa}.~A crude estimate using tree-level Madgraph~\cite{Alwall:2014hca} simulation finds that opening the lepton invariant mass cut from being just around the $Z$ pole to simply requiring $M_{\ell\ell} > 4$ GeV roughly doubles the background.~This should also give a reasonable estimate for the fake photon background because the underlying process is $Z^{(*)}/\gamma^{(*)}$ + jets, so the invariant mass distribution when a photon is replaced with a jet should be similar.~Ultimately, the background is smooth and rapidly falling in the center of mass energy, allowing for good background discrimination with a simple side-band analysis.~Therefore, we do not expect relaxing the cuts on the lepton pair invariant mass to be an obstruction for enhancing the virtual diphoton signal.

\section{Decays to four leptons}

We now turn to $\varphi \rightarrow 4\ell$ four body decays where again $4\ell = 2e2\mu, 4e, 4\mu$.~In this case the operator
\bea
\frac{c_{ZZ}}{4\Lambda}\, \varphi \,Z_{\mu\nu}\widetilde{Z}^{\mu\nu}
\eea
will also contribute and is naively the dominant effect due to the fact that both $Z$ bosons can be on-shell at $750$~GeV.~There are however, still contributions from the $c_{\gamma\gamma}$ and $c_{Z\gamma}$ operators studied in the previous section.~If these operators descend only from the $SU(2)\times U(1)$ invariant operators of~\eref{SM-tensor}, then there are only two unknowns and the system is over-constrained.~Therefore, measuring the contribution of all three operators is a non-trivial test of the SM gauge symmetry at the scale of the mass of the new resonance.~While the $\varphi \to 4\ell$ rate alone is not enough to measure all three operators, a fully differential analysis may be able to determine all three in a single channel~\cite{followup}, but we do not explore this here. 
 
The current best limits for decays to $ZZ$ in Run I come from the $\ell \bar\ell q\bar{q}$ channel~\cite{Aad:2015ipg} from which one can extract that the $\varphi$ decay to $ZZ$ is at most a factor of six bigger than the rate to $\gamma\gamma$~\cite{Franceschini:2015kwy} assuming that $\varphi$ is produced from gluon initial states.~This channel has a significantly higher branching ratio than the $4\ell$ channel, but suffers from a worse signal to background ratio.~Therefore, this search requires that both pairs of objects are roughly on the $Z$ pole.~There is also a search for decays to four leptons~\cite{Aad:2015rka} which has a significantly smaller rate, but is experimentally much cleaner.~In this search, there is also a requirement that one lepton pair invariant mass be between 50 and 120 GeV while the second is required to be between 12 and 120 GeV.~This not only reduces the total signal rate, but also the relative size of any non-$ZZ$ contribution to $\varphi \to 4\ell$, analogous to the three body case of $\varphi \to 2\ell\gamma$ described above. 
 
Here we will study ratios of partial widths involving $\varphi \rightarrow 4\ell$ in the two dimensional parameter space of $\lambda_{Z\gamma}$ defined in~\eref{lambdazg} and a second ratio of couplings,
\bea\label{eq:lamZZ}
\lambda_{ZZ} = c_{ZZ}/c_{\gamma\gamma}.
\eea
The kinematics of $\varphi \rightarrow 4\ell$ are more complicated than $2\ell\gamma$ and have been studied at length in the context of a heavy Higgs decay (see for example~\cite{Soni:1993jc,Barger:1993wt,Choi:2002jk,Buszello:2002uu,Gao:2010qx}).~Although there are multiple angular observables which contain useful information, in this simplified study we focus on the information contained in the two invariant mass distributions of the lepton pairs.~In particular, as with our study of $\varphi \to 2\ell\gamma$, we examine how the $\varphi \to 4\ell$ rate as well as its composition in terms of the $Z Z, Z \gamma^\ast$, and $\gamma^\ast\gamma^\ast$ components is affected by phase space cuts on the invariant mass of the lepton pairs.

We label the lepton pair invariant masses $M_1$ and $M_2$ and define $M_1 > M_2$ following the conventions and definitions in~\cite{Chen:2012jy,Chen:2013ejz}.~Since we are considering only rates, the difference between the $2e2\mu$ and $4e/4\mu$ channels due to identical final state interference is negligible.~However, as pointed out in~\cite{Chen:2015iha}, these identical final state effects can greatly influence event selection and these channels should be treated separately in a more complete fully differential likelihood analysis~\cite{Chen:2013ejz,Chen:2014gka,Chen:2014pia,Chen:2015iha}.~Since these subtleties are not relevant for current purposes, we simply study the $2e2\mu$ channel and multiply by a factor of two to include $4e$ and $4\mu$.

We first consider the ratio of the $\varphi\rightarrow 4\ell$ rate to the $\varphi\rightarrow \gamma\gamma$ decay rate as shown~\fref{4lrate}.~As with the Higgs boson at 125~GeV, this ratio will not be very large, but this is compensated by the very high precision with which it can be measured~\cite{Djouadi:2015aba}.~Depending on the coupling ratios, the $\varphi\rightarrow 4\ell$ rate will not be bigger than $\mathcal{O}(2-3\%)$ of the $\varphi\rightarrow \gamma\gamma$ rate for coupling ratios which are still allowed by $\varphi \rightarrow ZZ$ and $\varphi \rightarrow Z\gamma$ direct searches~\cite{Franceschini:2015kwy}.~This happens to be roughly the same as for the 125~GeV Higgs boson where this ratio is $\sim 2.5\%$~\cite{Dittmaier:2011ti,Heinemeyer:2013tqa}.~As the Higgs boson was discovered in both $h\to \gamma\gamma$ and $h\to 4\ell$~\cite{Chatrchyan:2012xdj,Aad:2012tfa}, this gives some hope that if the $750$~GeV diphoton excess persists, a signal in $\varphi \to 4\ell$ may also be observable soon.

From~\fref{4lrate}, we also see that the rate can be enhanced by going to more inclusive phase space cuts: $4 < M_{1,2} < 750$ GeV, compared to those used by the ATLAS search~\cite{Aad:2015ipg} which requires $50 < M_1 < 120$ GeV and $12 < M_2 < 120$~GeV.~The effect is largest when $\lambda_{ZZ} \ll 1$ since in this case the $Z\gamma^\ast$ and $\gamma^\ast\gamma^\ast$ components make up a larger fraction of $\varphi \to 4\ell$.~Thus phase space cuts have a larger effect compared to when $ZZ$ dominates, since in that case both $Z$ bosons can be on-shell in either the more inclusive or the ATLAS-like cuts.~We again show values of $\lambda_{ZZ}$ and $\lambda_{Z\gamma}$ larger than allowed by $\varphi \to ZZ$ and $\varphi \to Z\gamma$ searches~\cite{Franceschini:2015kwy} due to the various assumptions which go into these limits as discussed above. 
\begin{figure}[tb]
\includegraphics[width=.45\textwidth]{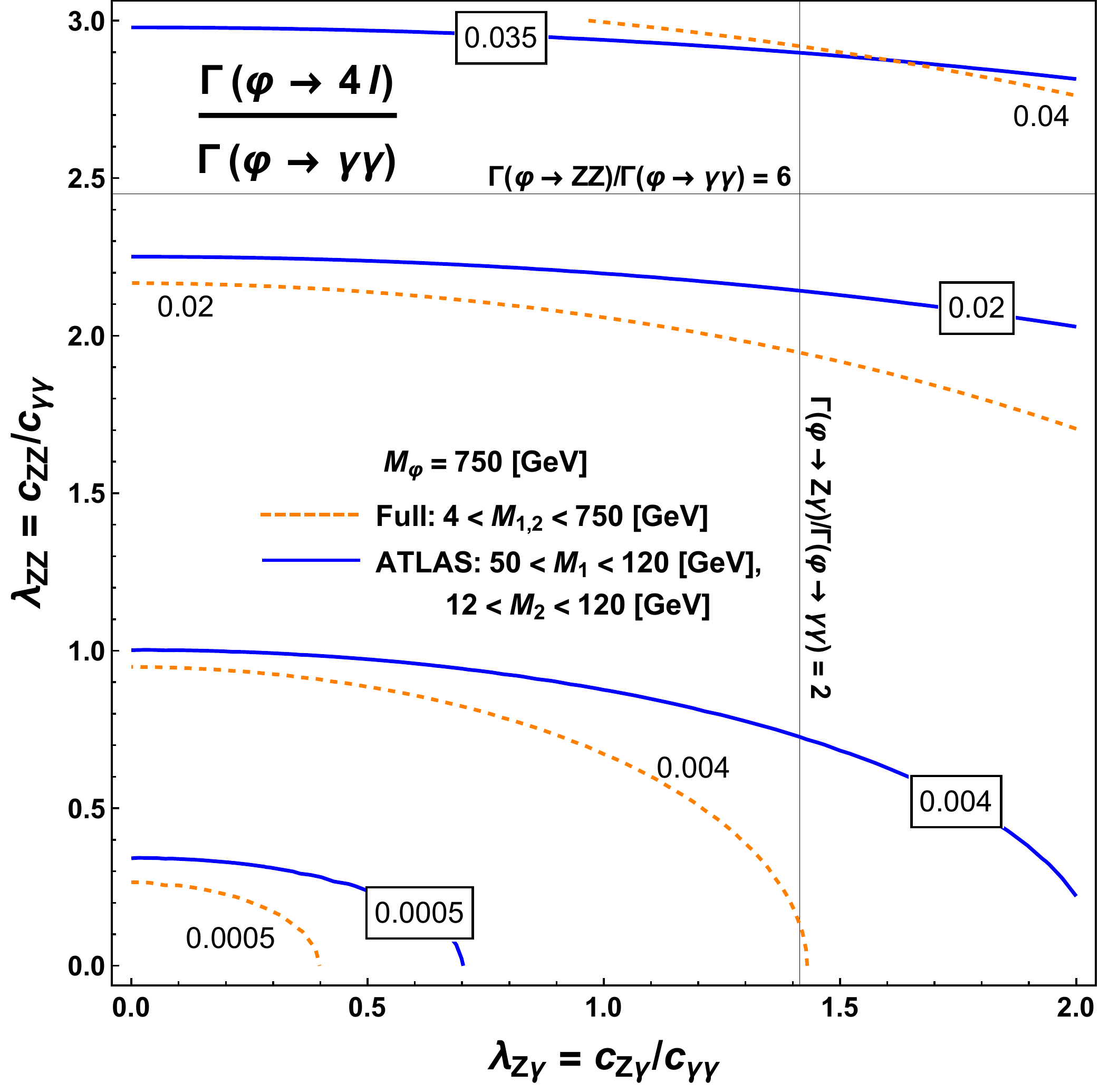}
\caption{Contours for the ratio of the rate of $\varphi\rightarrow 4\ell$ ($4\ell \equiv 2e2\mu + 4e + 4\mu$) over the rate of $\varphi\rightarrow \gamma\gamma$ in the plane of coupling ratios $\lambda_{Z\gamma}$ and $\lambda_{ZZ}$ defined in~\eref{lambdazg} and~\eref{lamZZ}, respectively.~We show inclusive phase space cuts $4 < M_{1,2} < 750$~GeV (dashed orange) as well as ATLAS-like cuts~\cite{Aad:2015ipg} with $50 < M_1 < 120$~GeV and $12 < M_2 < 120$~GeV (solid blue).~We also put the limits on the coupling ratios coming from $\varphi \to ZZ$ and $\varphi \to Z\gamma$ searches assuming $\varphi$ is produced from gluon initial states as in~\cite{Franceschini:2015kwy}.  }
\label{fig:4lrate}
\end{figure}

In~\fref{4lcomponents} we show the relative contribution of the naively subdominant components to $\varphi\rightarrow 4\ell$, namely those arising from $Z\gamma^\ast$ (blue) and $\gamma^\ast\gamma^\ast$ (orange).~Again we see that expanding the phase space cuts gives significantly more sensitivity to these components than current ATLAS cuts.~The absolute size of the $ZZ$ component is relatively unaffected when $\lambda_{ZZ} \gtrsim 1$ and $\lambda_{Z\gamma} \ll 1$ as can also be inferred from~\fref{4lrate} because the inclusive and ATLAS cut contours become very similar in that region.~We also see in~\fref{4lcomponents} that the $Z\gamma^\ast$ component dominates when $\lambda_{ZZ} \ll 1, \lambda_{Z\gamma} \gtrsim 1$, and that the $\gamma^\ast\gamma^\ast$ dominates when $\lambda_{ZZ} \ll 1, \lambda_{Z\gamma} \ll 1$.~Finally, we note the sharply rising slope for the size of the $\gamma^\ast\gamma^\ast$ component when $\lambda_{ZZ} \lesssim 0.3$ and $\lambda_{Z\gamma} \lesssim 0.5$, indicating a strong sensitivity in this regime.
\begin{figure}[tb]
\includegraphics[width=.45\textwidth]{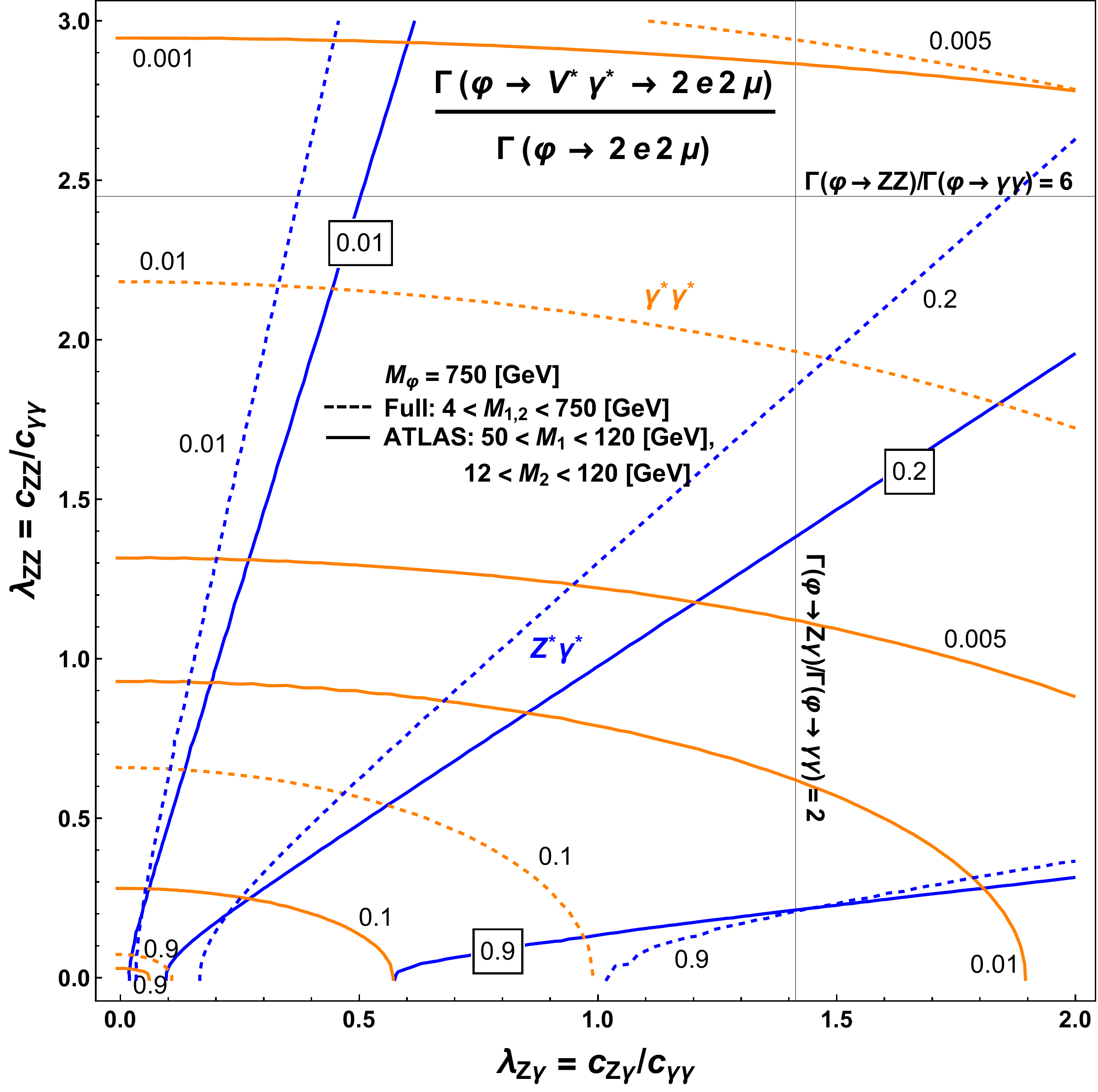}
\caption{Contours for the fraction of $2e2\mu$ events that come from $Z\gamma^\ast$ (blue) and $\gamma^\ast\gamma^\ast$ (orange) in the plane of coupling ratios $\lambda_{Z\gamma}$ and $\lambda_{ZZ}$ defined in~\eref{lambdazg} and~\eref{lamZZ}, respectively.~Again the dashed lines correspond to inclusive phase space cuts while the the solid lines correspond to the phase space cuts used in the ATLAS search~\cite{Aad:2015ipg} as defined in figure.~We also show limits on the coupling ratios assuming $\varphi$ is produced from gluon initial states as in~\cite{Franceschini:2015kwy}.  }
\label{fig:4lcomponents}
\end{figure}

We again propose a simple way to measure $\lambda_{Z\gamma}$ and $\lambda_{ZZ}$ analogous to the one from the previous section for $\varphi \to 2\ell\gamma$.~Namely we define a similar ratio
\bea
R_{ZZ}(\Delta_{i}) = \frac{N(M_Z + \Delta_{1,2} > M_{1,2} > M_Z - \Delta_{1,2}) }{{\rm total \; number \; of \; events}}.
\label{eq:rzz}
\eea
where again the total number of events is defined by the inclusive phase space with $4 < M_{1,2} < 750$~GeV.~We show contours of $R_{ZZ}$ in~\fref{rzz} where we see that it is very sensitive to 
$\lambda_{ZZ}$ for $\lambda_{ZZ} \lesssim 1$ while less sensitive to $\lambda_{Z\gamma}$.~The stronger sensitivity to $\lambda_{ZZ}$ can be understood from the pole structure of the two $Z$ bosons which can both be on-shell at 750~GeV.~Again we also see the benefits of using more inclusive cuts to enhance the non-ZZ components.
\begin{figure}[tb]
\includegraphics[width=.45\textwidth]{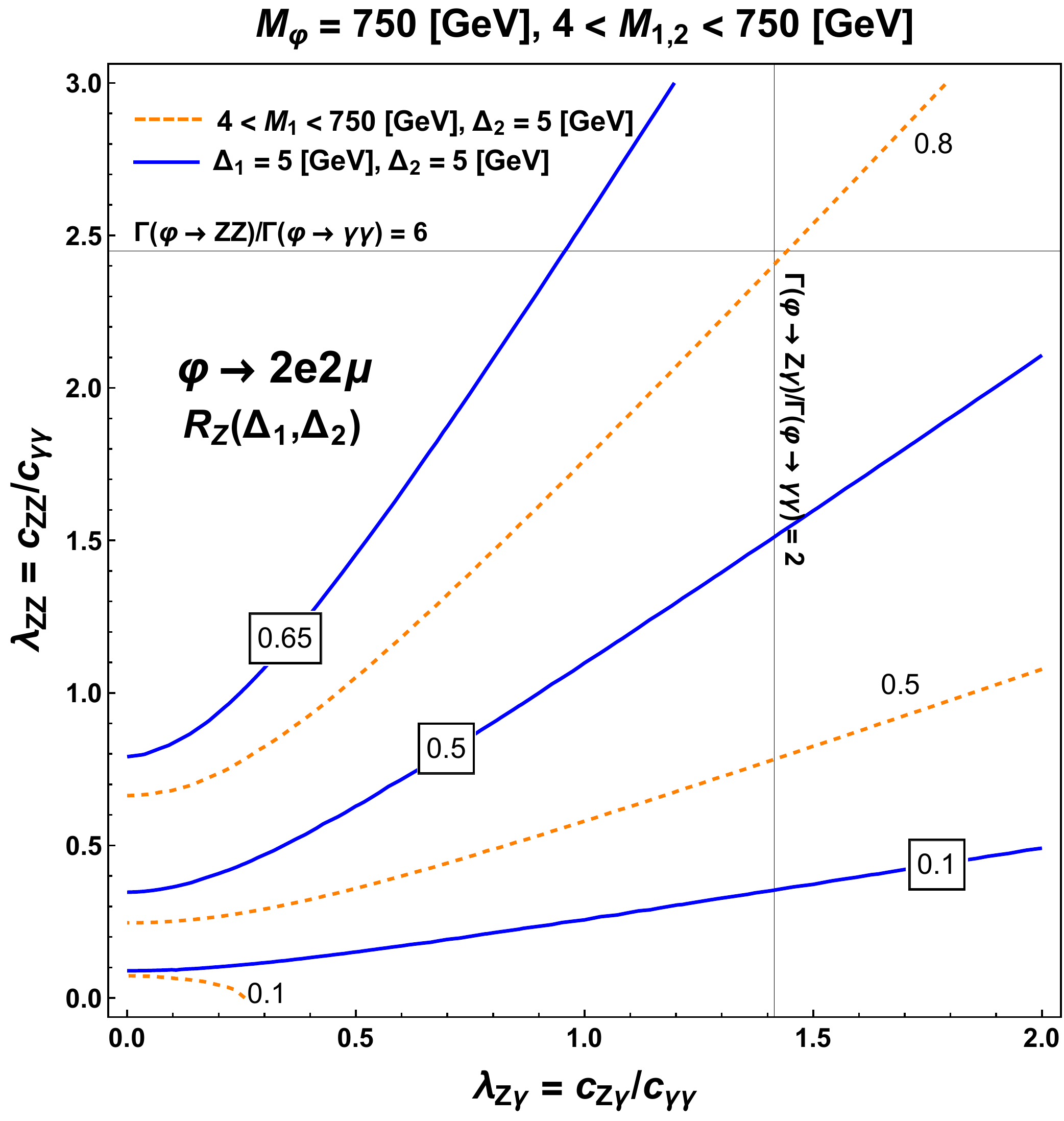}
\caption{Contours of $R_{ZZ}$ as defined in~\eref{rzz} in the plane of coupling ratios $\lambda_{Z\gamma}$ and $\lambda_{ZZ}$ defined in~\eref{lambdazg} and~\eref{lamZZ} respectively.~We show $R_{ZZ}$ for $\Delta_1=\Delta_2=5$ GeV (blue solid) and $\Delta_2 = 5$ GeV with $4 < M_{1} < 750$~GeV (dashed orange).~We also show the limits on the coupling ratios assuming $\varphi$ is produced from gluon initial states as in~\cite{Franceschini:2015kwy}.  }
\label{fig:rzz}
\end{figure}

We have not discussed interference between the different intermediate states since it has a negligible effect on the rates.~However, in a fully differential analysis where shape information is used, these interference effects can potentially be important.~In particular, as has been shown in many studies of the Higgs boson, these interference effects would give us access to the CP properties of $\varphi$ and to potential CP violating effects.~An investigation of these interesting possibilities using the framework of~\cite{Chen:2012jy,Chen:2013ejz,Chen:2014ona,Chen:2014pia} is ongoing~\cite{followup}.

In an experimental analysis, backgrounds must of course be taken into account, but, as with Higgs decays to four leptons, the background is very small.~The dominant source of background is quark initiated $ZZ$ production, with the one-loop gluon process also contributing, but again is very subdominant~\cite{Aad:2015rka}.~As far as we know, there are no higher order calculations of these backgrounds, but that is partially because they are quite small.~As with the $2\ell\gamma$ case, enlarging the mass window from the current searches will increase the background, but it will still be small and smooth, so a sideband analyses can again be used.~Of course using a fully differential likelihood analysis would increase the ability to discriminate signal from background further~\cite{Gainer:2011xz}, but we do not investigate this possibility here.~For present purposes we have simply used naive estimates to ensure that the dominant background can easily be controlled.

\section{Non-leptonic and Exotic Decays}

If the diphoton excess proves to be more than a statistical fluctuation and indeed due to a new scalar $\varphi$, we will want to search for $\varphi$ decays in as many channels as possible, not just the experimentally clean ones with leptons.~Furthermore, our considerations of the virtual diphoton contributions to $\varphi \to 2\ell\gamma$ and $\varphi \to 4\ell$ also apply when one considers other charged fermions in the SM, though of course experimentally these channels are much less cleanly measured.~While the branching ratios and couplings of the $Z$ and photon are well measured, looking for decays in $\varphi \to 2f\gamma$ and $\varphi \to 2f2f'$ is an important test to see if there is other new physics or couplings of $\varphi$ to SM fermions.
 
In~\fref{ffgam} we consider the $\varphi \to V\gamma \to 2f\gamma$ (where $V = Z, \gamma^\ast$) partial width normalized to $\varphi \rightarrow \gamma\gamma$ for the various light SM fermions.~We see that for small $\lambda_{Z\gamma}$ the leptons (solid red) dominate.~By comparing the solid red curve, which is the full $\varphi \to 2\ell \gamma$ decay width, to the dashed red curve which is only the on-shell $Z\gamma$ mediated $\varphi \to 2\ell\gamma$ width, we see that the low $\lambda_{Z\gamma}$ behavior is dominated by photon contributions.~This explains why the leptons are the largest contribution at small $\lambda_{Z\gamma}$, since they have larger electric charge than SM quarks.~At larger $\lambda_{Z\gamma}$, decays with quarks and neutrinos become more important.~While these are experimentally more difficult, kinematic shape information can perhaps be used to uncover the signal from the background, though we do not explore this issue here.~We also note that if one imposes the $\varphi \to Z\gamma$ limit (vertical line) derived from~\cite{Franceschini:2015kwy}, this implies a limit on $\varphi \to 2\ell\gamma$ ($\ell = e, \mu, \tau$) and $\varphi \to 2q\gamma$ ($q = u, d, c, s, b$) of $\sim 40\%$ and $\sim 90\%$ of the $\gamma\gamma$ rate, respectively.~ Of course, all the caveats discussed above about the production mechanism still apply.

\begin{figure}[tb]
\includegraphics[width=.45\textwidth]{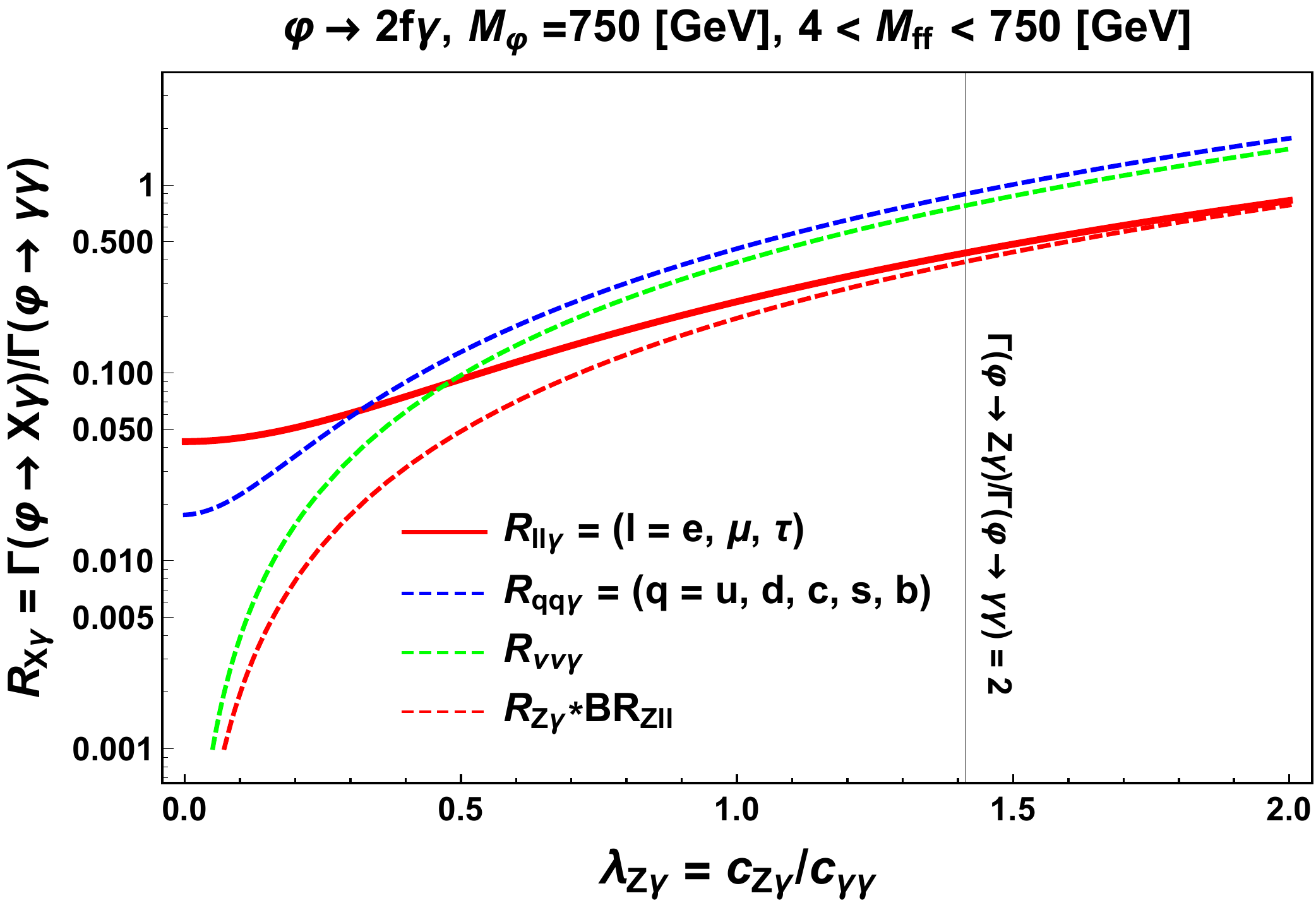}
\caption{The ratio of the $\varphi \to 2f\gamma$ partial width relative to $\varphi \to \gamma\gamma$.~This is plotted as a function of the ratio $\lambda_{Z\gamma}$ defined in~\eref{lambdazg}.~We plot $f$ to be leptons (solid red), light quarks (dashed blue), and neutrinos (dashed green).~The on-shell $Z\gamma$ contribution times branching ratio into leptons (dashed red) is shown for comparison.~We also show the limit (vertical line) obtained from $Z\gamma$ searches~\cite{Franceschini:2015kwy} at 8 TeV. }
\label{fig:ffgam}
\end{figure}

For the case of four fermion decays, there are many more possibilities including $\ell\ell\nu\nu$ and $\ell\ell q \bar{q}$ which are also experimentally challenging.~They are expected however to have much larger rates than $\varphi \to 4\ell$ in much of the parameter space, particularly when the $\varphi\gamma\gamma$ coupling is not parametrically larger than for $Z\gamma$ and $ZZ$.~One can also consider $WW$ decays to $\ell\nu\ell\nu$ or other channels.~While the computations utilized in this work can be extended to these cases as well, the experimental analyses become more difficult and backgrounds have to be treated more carefully.~If the resonance at 750 GeV turns out to be genuine new physics, fully understanding all these channels will be crucial to characterizing the new state and any theory it might be associated with. 

Finally, we note that the simplified analysis presented here is also useful if the new physics is not one simple resonance decaying to diphoton but is instead multiple resonances~\cite{Franceschini:2015kwy,Potter:2016psi}, not a resonance~\cite{Cho:2015nxy,Li:2015jwd,An:2015cgp,Bernon:2015abk,Liu:2015yec}, or a resonance that decays through a cascade~\cite{Knapen:2015dap,Falkowski:2015swt,Agrawal:2015dbf,Chang:2015sdy,Chala:2015cev}.~Each of these kinds of models has different predictions for both the correlated searches via $Z\gamma$ and $ZZ$ as well as with virtual photons.~Furthermore, the improved signal to background ratio, particularly in the case of four leptons, will allow a more precise measurement of the line-shape allowing discrimination of many possibilities.~Should the excess persist, an exploration of these cases would also be interesting. 

\section{Conclusions and Outlook}
\label{sec:conclution}

In this work, we interpret the excess observed by ATLAS and CMS in the diphoton spectra around 750 GeV to be indicative of a new scalar resonance $\varphi$ decaying to photons.~We show in particular that the effective operator responsible for the $\varphi \to \gamma\gamma$ decay will also lead to a signal in $\varphi \to 2f\gamma$ and $\varphi \to 2f2f'$ (where $f$ is a SM fermion) decays independently of the effective couplings of $\varphi$ to $Z\gamma$ and $ZZ$.~We have focused in particular on the leptonic $\varphi \to 2\ell\gamma$ and $\varphi \to 4\ell$ channels ($\ell = e, \mu$).~Depending on the relative sizes of these effective couplings, we show that the virtual diphoton component can make up a sizable, and sometimes dominant, contribution to the total $\varphi \to 2\ell \gamma$ and $\varphi \to 4\ell$ partial widths. 

We have also explored the effects that phase space cuts on the invariant mass of the lepton pairs have on the total rates and composition of $\varphi \to 2\ell\gamma$ and $\varphi \to 4\ell$.~We have emphasized the contribution from virtual photons and pointed out that current experimental searches should be modified in order to enhance the sensitivity to these virtual photon effects.~We find that a more inclusive phase space cut (while still requiring the full system to be at the resonance mass) would allow an increased signal rate and larger contributions from all components of $\varphi \to 2\ell \gamma$ and $\varphi \to 4\ell$.~The virtual photon contributions in particular can be increased by an order of magnitude.~This allows us to study $\varphi$ in more detail while still keeping the backgrounds under control.

Finally, we have used a simple cut and count method with ratios of partial widths to assess the potential sensitivity of $\varphi \to 2\ell\gamma$ and $\varphi \to 4\ell$ to ratios of effective couplings between $\varphi$ and $ZZ$, $Z\gamma$, and $\gamma\gamma$.~We find particularly strong sensitivity when the effective coupling of $\varphi$ to $\gamma\gamma$ is larger than to $Z\gamma$ or $ZZ$.~A full analysis taking advantage of all the final state kinematics can reveal more about the nature of the resonance, but we have left this to ongoing work.~We have also briefly discussed non-leptonic channels and potential applications of our analysis methods to more exotic possibilities for explaining the diphoton excess.~Should the excess persist, the methods utilized and discussed here will prove useful for ascertaining the nature of the putative new resonance. 

\vspace{.3in}

{\bf Acknowledgments:}~We would like to thank Yi Chen, Jose Santiago, and Jorge de Blas for useful conversations.~R.V.M.~is supported by MINECO, under grant number FPA2013-47836-C3-2-P. 
 
\bibliographystyle{apsrev}
\bibliography{References}

\end{document}